\documentclass{JINST}

\title{Performance of a New Electron-Tracking Compton Camera under Intense Radiations from a Water Target irradiated with a Proton Beam}

\author{
	Y.~Matsuoka$^a$\thanks{Corresponding author.},
	T.~Tanimori$^a$,  
	H.~Kubo$^a$, 
	A.~Takada$^a$,
	J.~D.~Parker$^a$, 
	T.~Mizumoto$^a$, 
	Y.~Mizumura$^{a,b}$,
	S.~Iwaki$^a$, 
	T.~Sawano$^a$,   
	S.~Komura$^a$, 
	T.~Kishimoto$^a$,
	M.~Oda$^a$,
	T.~Takemura$^a$,
	S.~Miyamoto$^a$,
	S.~Sonoda$^c$, 
	D.~Tomono$^a$, 
	K.~Miuchi$^d$, 
	S.~Kabuki$^e$,
	and S.~Kurosawa$^f$\\
	\llap{$^a$}Division of Physics and Astronomy, Graduate School of Science, Kyoto Univ.,
		Kyoto, Kyoto, 606-8502, Japan\\
	\llap{$^b$}Unit of Synergetic Studies for Space, Kyoto Univ.,
		Kyoto, Kyoto, 606-8502, Japan\\
	\llap{$^c$}Advanced Biomedical Engineering Research Unit, Kyoto Univ.,
		Kyoto, Kyoto, 606-8502, Japan\\
	\llap{$^d$}Department of Physics, Graduate School of Science, Kobe Univ.,
		Kobe, Hyogo, 657-8501, Japan\\
	\llap{$^e$}Department of Radiation Oncology, Tokai Univ.,
		Isehara, Kanagawa, 259-1193, Japan\\
	\llap{$^f$}Institute for Materials Research, Tohoku Univ.,
		Sendai, Miyagi, 980-8577, Japan\\
	E-mail: \email{matsuoka@cr.scphys.kyoto-u.ac.jp}
}

\abstract{
We have developed an electron-tracking Compton camera (ETCC)
for use in next-generation MeV gamma ray telescopes.
An ETCC consists of a gaseous time projection chamber (TPC) and pixel scintillator arrays (PSAs). 
Since the TPC measures the three dimensional tracks of Compton-recoil electrons, 
the ETCC can completely reconstruct the incident gamma rays. 
Moreover, the ETCC demonstrates efficient background rejection power in Compton-kinematics tests, 
identifies particle from the energy deposit rate (dE/dX) registered in the TPC, and provides high quality imaging by completely reconstructing the Compton scattering process. 
We are planning 
the ''Sub-MeV gamma ray Imaging Loaded-on-balloon Experiment'' (SMILE)
for our proposed all-sky survey satellite. 
Performance tests of a mid-sized (30 cm)$^{3}$ ETCC, constructed for observing the Crab nebula, are ongoing. 
However, observations at balloon altitudes or satellite orbits are obstructed by radiation background from the atmosphere and the detector itself \cite{takada_APJ_2011}.
The background rejection power was checked using proton accelerator experiments conducted
at the Research Center for Nuclear Physics, Osaka University. 
To create the intense radiation fields encountered in space, which comprise gamma rays, neutrons, protons, and other energetic entities, we irradiated a water target with a 140~MeV proton beam and placed a SMILE-II ETCC near the target.
In this situation, the counting rate was five times than that expected at the balloon altitude.
Nonetheless, the ETCC stably operated and identified particles sufficiently to obtain a clear gamma ray image of the checking source. 
Here, we report the performance of our detector and demonstrate its effective background rejection based in electron tracking experiments. 
}


\begin{document}

\section{Introduction}
Gamma rays with energies of 100~keV to 1~MeV are expected to provide valuable insight into unexplained celestial phenomena,
such as nucleosynthesis in supernovae \cite{SNR_ref},
particle acceleration in active galactic nuclei \cite{AGN_ref}
and gamma ray bursts \cite{GRB_ref}.
Therefore, gamma rays in this energy band are of immense interest to astronomers.
However, such gamma rays are difficult to observe
because it is needed to lunch at balloon altitudes or satellite orbits
since an atmosphere shields the gamma rays \cite{ref_atmosphere}
and the detector and its surrounding materials are strongly intercepted by cosmic rays.
To date, the COMPTEL onboard the Compton Gamma Ray Observatory has observed celestial objects in the MeV gamma ray band. COMPTEL rejects backgrounds with significant time of flight (TOF) between the first and second detector \cite{COMPTEL_TOF}.
However, because this criterion rejections an insufficiemt number of objects in an orbit background \cite{COMPTEL_final},
 COMPTEL detects  several tens of objects \cite{COMPTEL_catalogue}.
 On the other hand, \emph{Fermi}-LAT found approximately two thousand sources in the sub-GeV/GeV gamma ray band during the first two~years of its all-sky survey \cite{Fermi_catalogue}.

\begin{figure}[htbp]
\centering
\includegraphics[height=.37\textwidth]{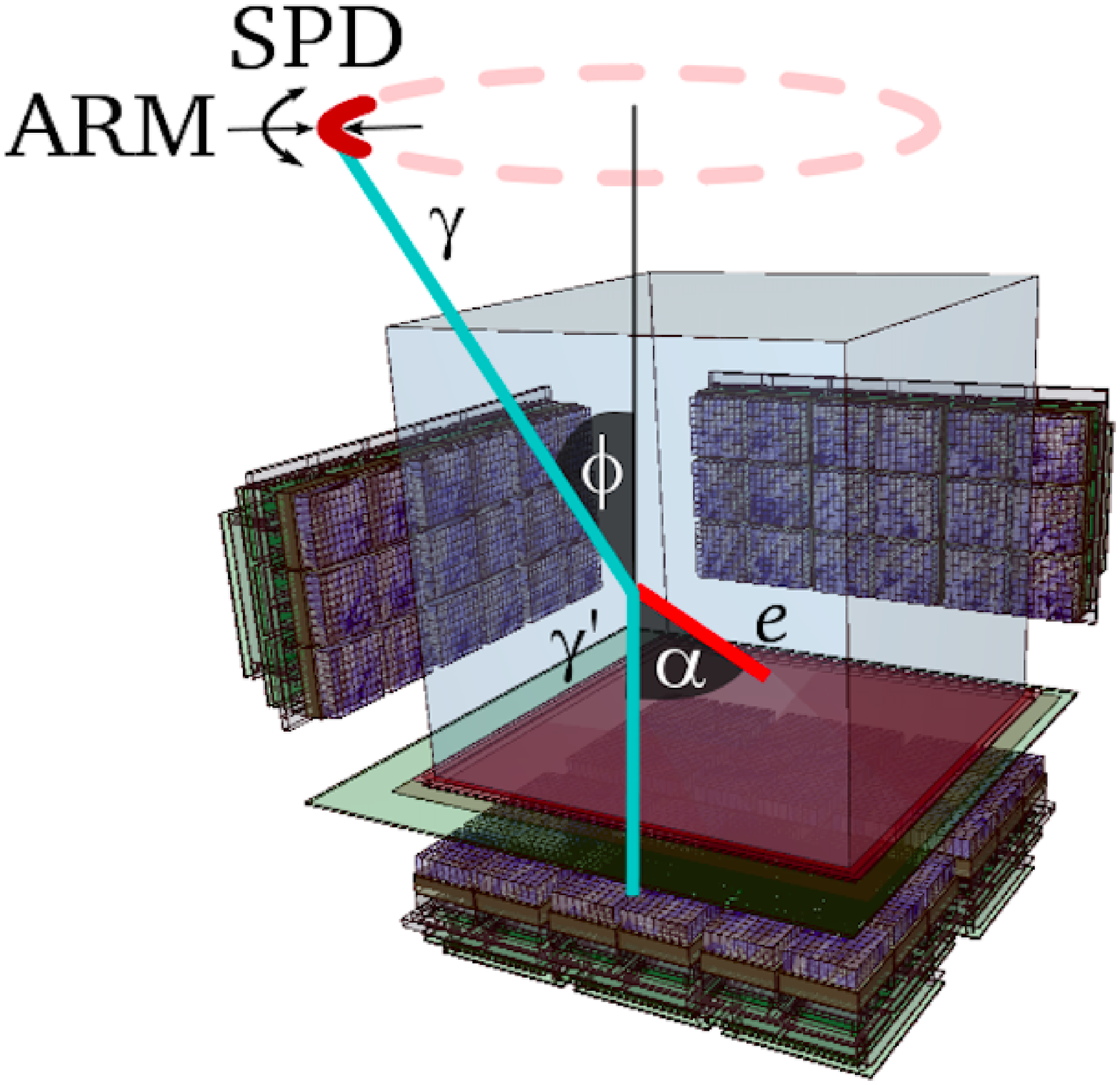}
\includegraphics[height=.37\textwidth]{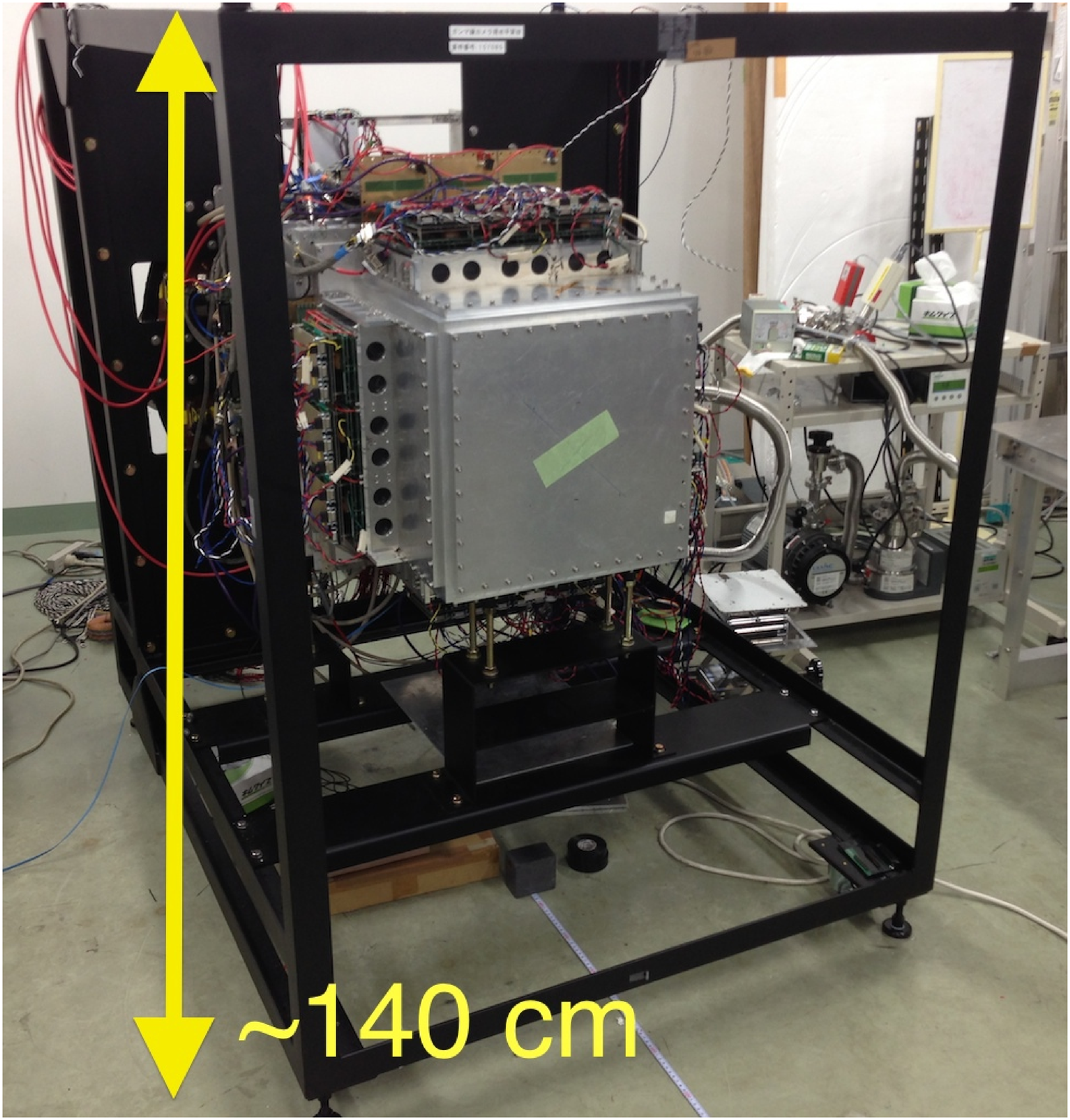}
\caption{
(\emph{Left panel}) Schematic of an ETCC. The ETCC consists of a gaseous electron tracker and scintillators, which firm a scattered gamma ray absorber.
An incident gamma ray, scattered gamma ray, recoil electron are labeled as $\gamma$, $\gamma$', and $e$, respectively. \newline
(\emph{Right panel}) Photograph of the SMILE-I\hspace{-.1em}I ETCC with jigs to see the horizontal direction.
}
\label{fig:ETCC}
\end{figure}

Given this background, we have developed an electron-tracking Compton camera (ETCC) for use in a next generation MeV gamma ray telescope (figure \ref{fig:ETCC}).
The ETCC consists of a gaseous time projection chamber (TPC) and pixel scintillator arrays (PSAs).
The TPC measures the energy of Compton recoil electrons and tracks their progress, 
while the PSAs measure the energies and positions of the scattered gamma rays.
Our gaseous TPC can measure the three-dimensional track of Compton recoil electrons to precise positional resolution. 
From this track information, we can deduce the direction and track length of the recoil electrons, parameters that cannot be obtained by usual Compton cameras.
Because the plane of the Compton scattering depends on the direction of the recoil electrons, 
the ETCC can determine not only the angular resolution measure (ARM), but also the scatter plane deviation (SPD).
Moreover, the angle $\alpha$ between the recoil electrons and scattered gamma rays can be geometrically obtained as 
\begin{equation}
	\cos \alpha_{geo} = {\bf g} \cdot {\bf e} ,
\end{equation}
where $\alpha_{geo}$ donates the geometrically obtained $\alpha$,
and $\bf g$ and $\bf e$ are unit vectors in the directions of the scattered gamma ray and recoil electron, respectively.
The kinematically calculated $\alpha$ ($\alpha_{kin}$) is given by
\begin{equation}
	\cos \alpha_{kin} = \left( 1 - \frac{m_e c^2}{E_\gamma} \right)
		\sqrt{\frac{K_e}{K_e + 2 m_e c^2}} ,
\end{equation}
where $E_\gamma$, $K_e$, $m_e$, and $c$ are the energy of the scattered gamma ray, the kinetic energy of the recoil electron, the electron mass, and the speed of light, respectively.
By comparing $\alpha_{geo}$ and $\alpha_{kin}$, our ETCC can reject random coincidence or multiple Compton scattering backgrounds.
From the energy deposit rate (dE/dX) in the TPC, we can identify the types of particles, intercepting the TPC.
Particle identification is enabled by the capacity of ETCC to reject neutron-recoil protons, incident charged particles, and electrons escaping from the TPC.
Together with Compton kinematic testing, this feature imbues the ETCC with powerful background rejection ability.

In 2006, we carried out the first balloon experiment ''Sub-MeV gamma ray Imaging Loaded-on-balloon Experimen'' (SMILE-I), using a small ETCC (installed with a 10~$\times$~10~$\times$~15~cm$^3$ TPC and 33~PSAs, effective area $\sim$0.01~cm$^2$) \cite{takada_APJ_2011}.
In this experiment, diffuse cosmic and atmospheric gamma rays were detected with efficient background rejection.
A milestone in the all-sky survey was the increase of the TPC size to (30 cm)$^3$ and the number of PSAs to 36,
which significantly improved the sensitivity of surveying \cite{ueno_JINST_2012}. 
Next, we constructed the balloon flight model ETCC (SMILE-I\hspace{-.1em}I FM),  
in which the detection capability of ETCC was assessed in 5$\sigma$, 4-h observations of bright celestial gamma ray sources such as the Crab Nebula and Cygnus X-1.
SMILE-I\hspace{-.1em}I requires an effective exceeding 0.5~cm$^2$ at 300~keV, and an ARM within 10$^{\circ}$ at 662~keV \cite{tanimori_SPIE_2012}.

We have checked the background rejection ability of the ETCC in huge backgrounds,
similar to those encountered at balloon altitudes or in outer space.
These backgrounds were emulated by a 140~MeV proton beam emitted from the cyclotron at the Research Center for Nuclear Physics (RCNP).
The present paper reports the performances of the SMILE-I\hspace{-.1em}I FM ETCC 
and the results obtained in the intense gamma ray and neutron radiation field generated by the proton beam.

\section{Instruments}
\subsection{Energy Resolutions of TPC and PSAs}
On Compton Cameras, the angular resolution of the telescope and the signal to noise ratio (S/N) depend on the ARM and SPD, respectively.
The ARM is determined by the precisions of the scattering angle ($\phi$) and scattering direction.
The scattering angle is calculated from the energies of the recoil electron and the scattered gamma ray.
Therefore, the energy resolution must be finely calibrated. 

The SMILE-I\hspace{-.1em}I FM ETCC consists of two detectors. 
The first is a (30~cm)$^3$ TPC based on (30~cm)$^2$ micro-pixel chamber ($\mu$PIC \cite{takada_NIMA_2007}), and a (31~$\times$~32)~cm$^2$ gas electron multiplier (GEM \cite{GEM_1997}). The gas is Ar:CF$_4$:iso-C$_4$H$_{10}$ (molar ratio 95:3:2), 1~atom.
The second detector comprises 108~PSAs with 8~$\times$~8~pixels of (6~$\times$~6~$\times$~13)~mm$^3$ GSO crystals, giving a total of 6912~pixels.

\begin{figure}[htbp]
\centering
\includegraphics[height=.35\textwidth]{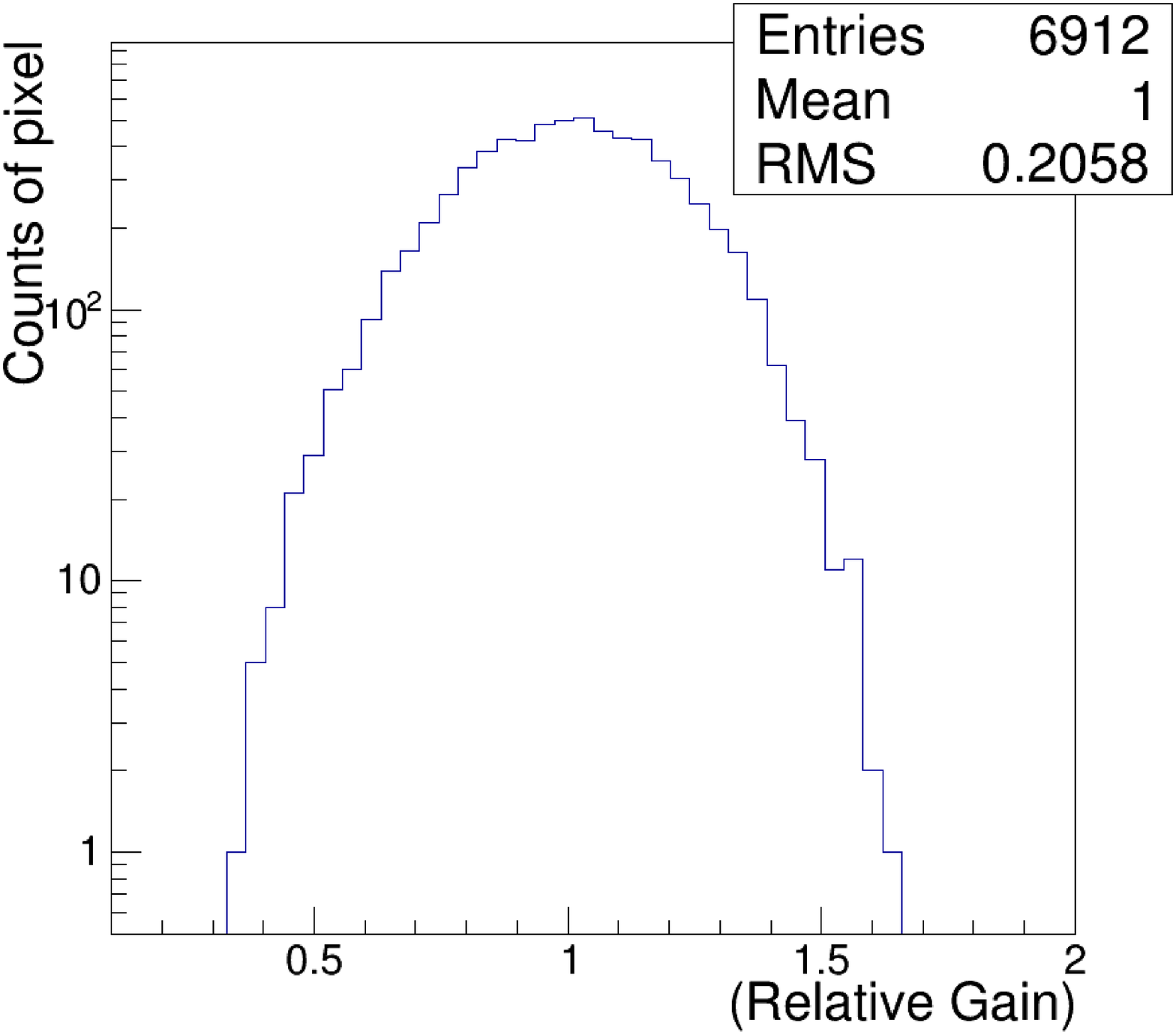}
\includegraphics[height=.35\textwidth]{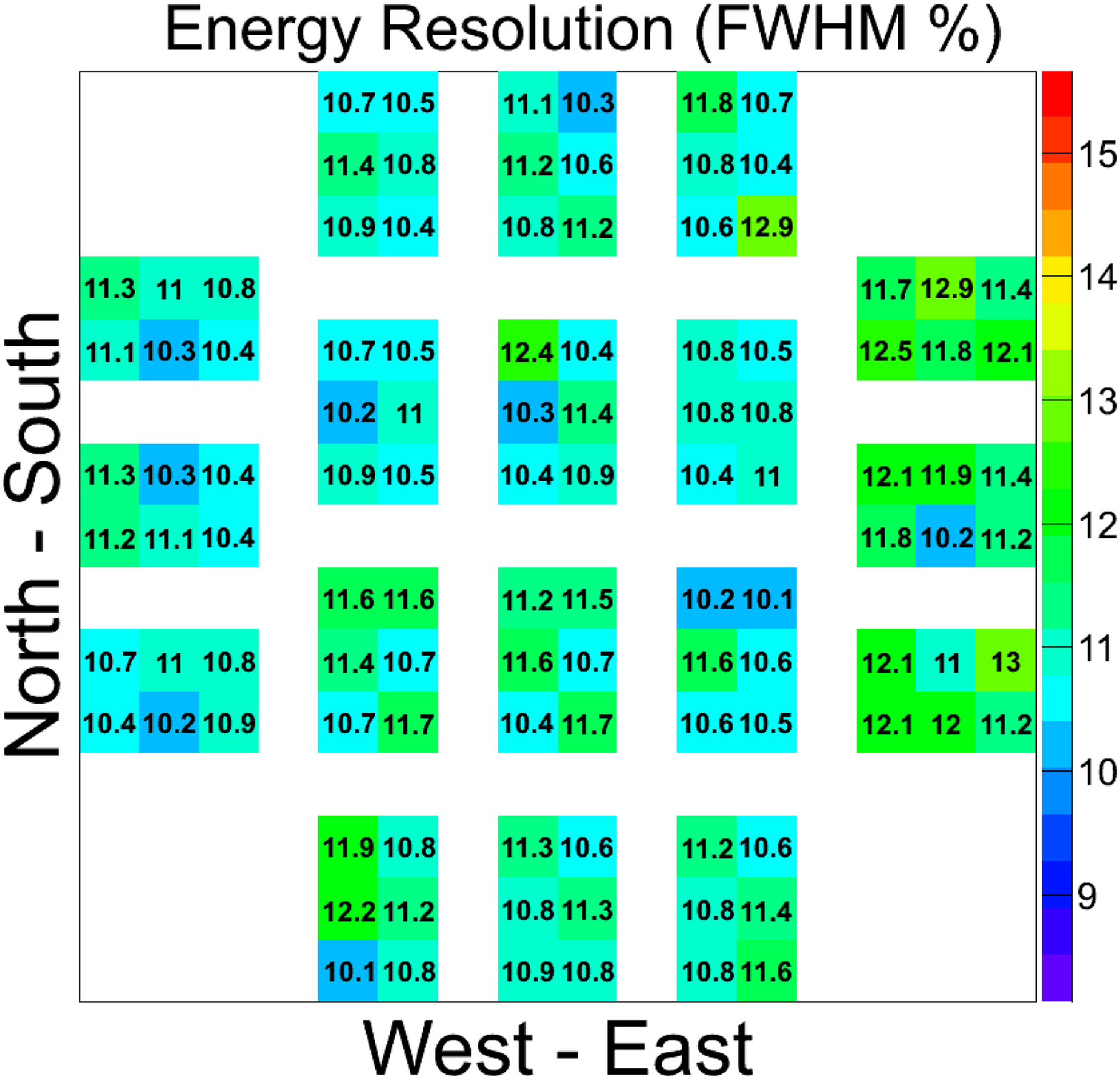}
\caption{
(\emph{Left panel}) Relative gain distribution of the 6912~GSO pixels in a PSA. \newline
(\emph{Right panel}) Energy resolution map of the PSAs at 662~keV.
The ETCC contains 36 bottom PSAs (central units) and (18~$\times$~4) side PSAs (outer units).
}
\label{fig:HA_Gain}
\end{figure}

The number of readout channels on each PSA can be reduced from 64~to~4 using a resistor-chain matrix.
The 64~pixels can then be clearly resolved by the center of gravity \cite{ueno_JINST_2012}. 
For each PSA, we measured the gains of all pixels and the total energy resolution.
The left panel of figure \ref{fig:HA_Gain} shows the relative gain distribution.
All pixels can detect gamma rays within the energy range $\sim$100~keV to $\sim$1~MeV.
The right panel is the energy resolution map of all PSAs;
the 36 central PSAs are placed at the bottom of the TPC 
and 18~PSAs are placed at each of the four sides.
The mean energy resolutionat 662~keV is 11.1~$\pm$~0.6~$\%$ FWHM.

\begin{figure}[!h]
\centering
\includegraphics[width=.4\textwidth]{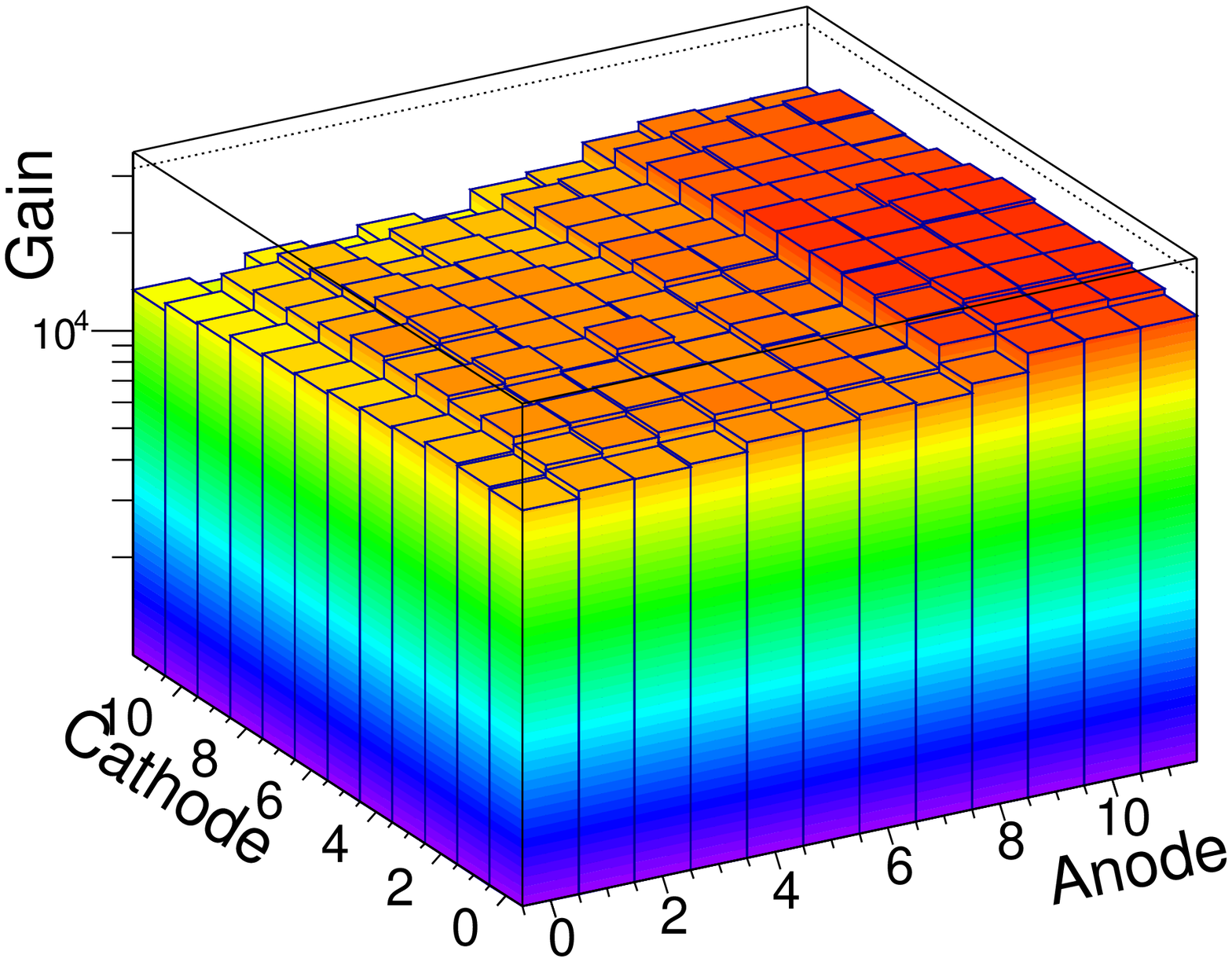}
\includegraphics[width=.4\textwidth]{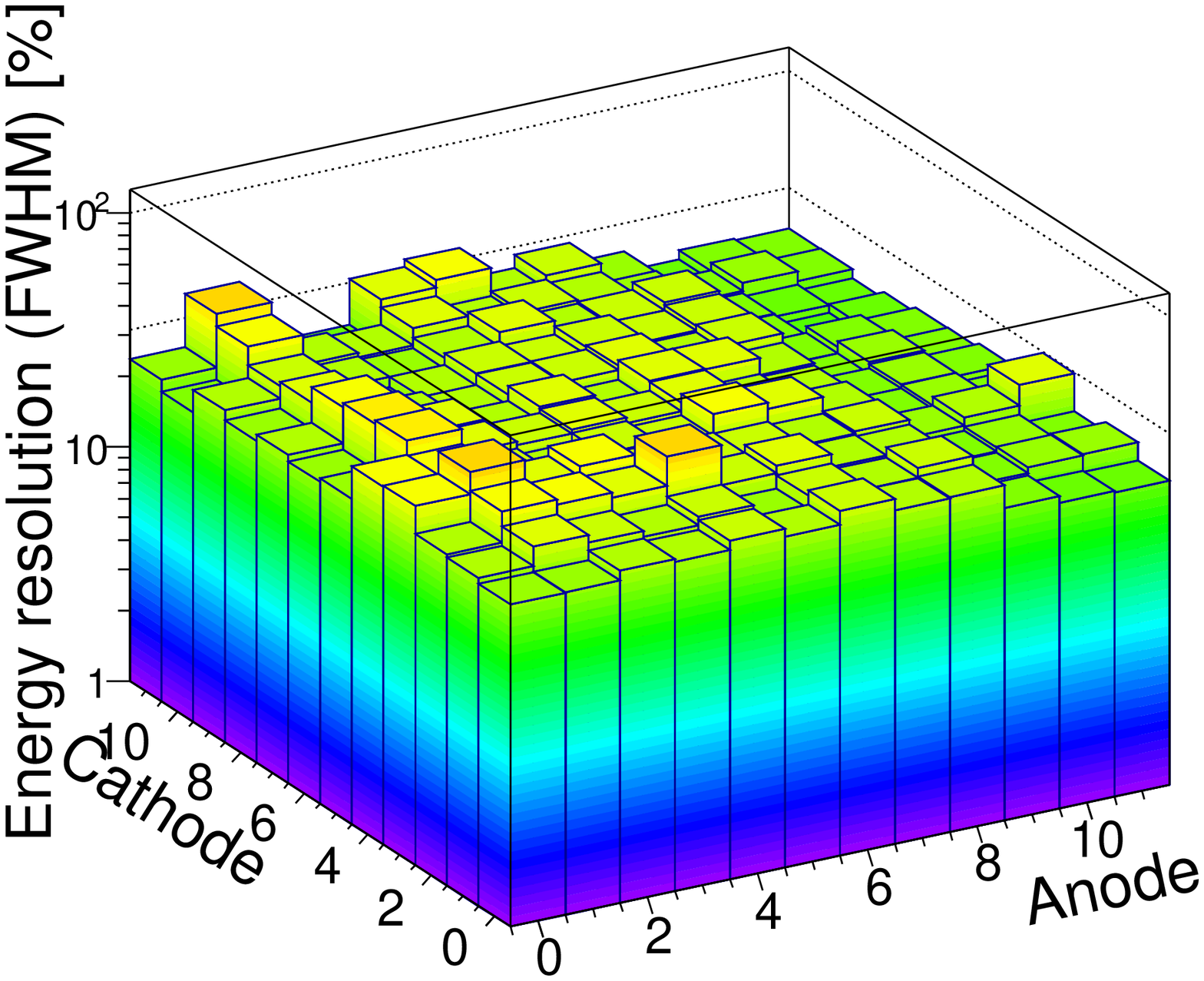}\\
\includegraphics[height=.28\textwidth]{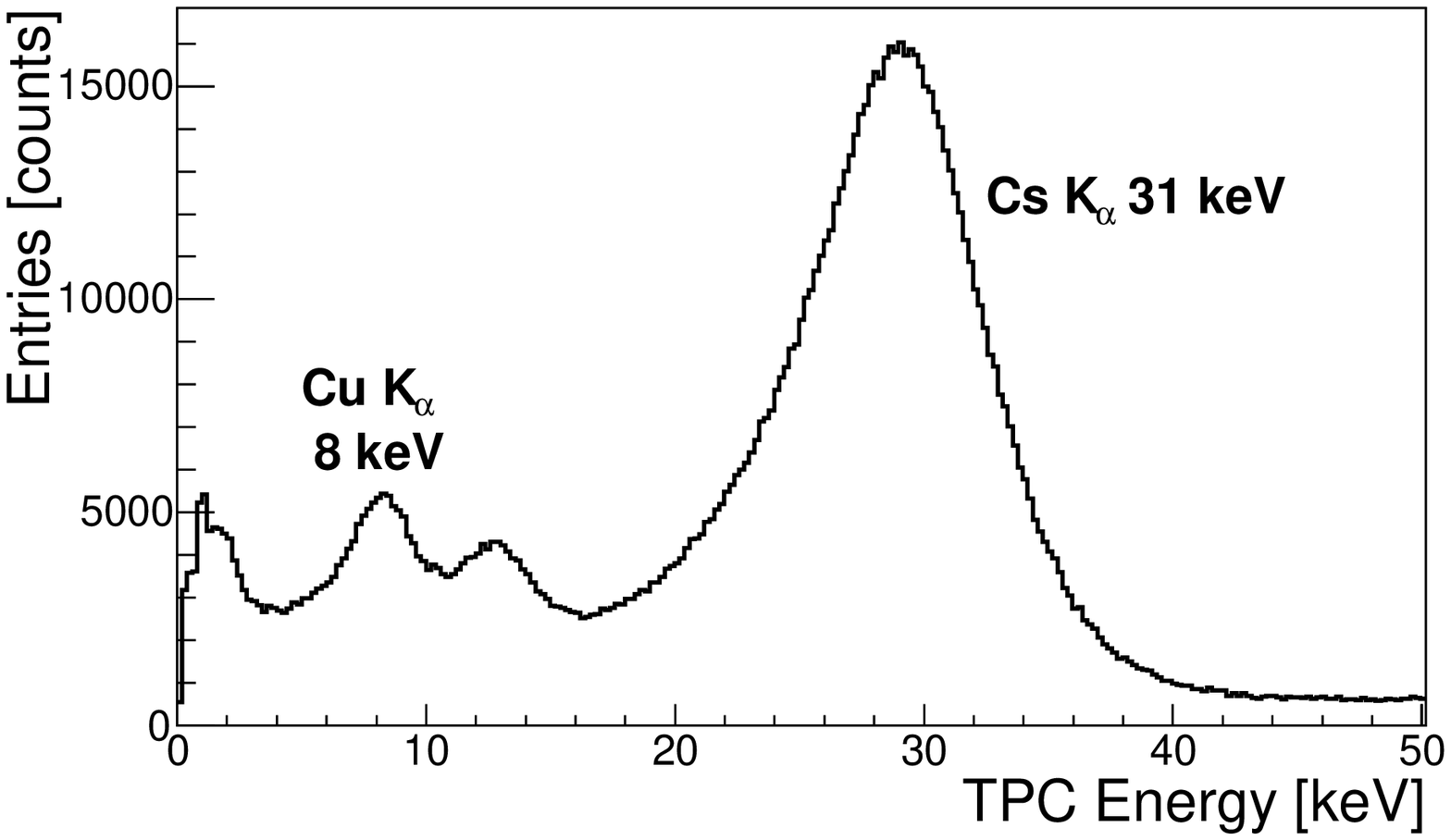}
\caption{
Gain map of the gaseous TPC of SMILE-I\hspace{-.1em}I ETCC divided into 12~$\times$~12 areas (upper left).
Distribution of energy resolution (FWHM) in each area (upper right).
$^{133}$Ba spectrum over the whole area (bottom).
}
\label{fig:TPC}
\end{figure}

The detection area of this $\mu$PIC is (30~$\times$~30)~cm$^2$.
There are 768~$\times$~768 pixels in the whole area, and each pitched at 400~$\mu$m.
The TPC signals are read by 768~anode strips and 768~cathode strips.
Our newly designed readout board \cite{mizumoto_IEEE_2013} measures the tracks of the charged particles as the time-over-threshold (TOT) per 2~strips and sums of the analog signals over 64~strips by Flash ADCs (FADCs).
For measuring the gas gain and the energy resolution, the $\mu$PIC was divided into 12~$\times$~12 equal areas.
The left panel of figure \ref{fig:TPC} shows the distribution of gains.
The mean and uniformity of the gas gain is $\sim$~2~$\times$~10$^4$ and 15.9~$\%$ RMS, respectively.
The energy resolution was measured using the 31~keV X-ray of Cs~K$_{\alpha}$ emitted from $^{133}$Ba.
The bottom panel of figure \ref{fig:TPC} shows the energy resolution across the whole area (24.8~$\%$ FWHM), and
the right panel shows the energy resolution distribution (18~$\%$ - 40~$\%$ FWHM).

\subsection{Data Suppression of TPC to ensure High Event Rate}
To reduce the power consumption and the dead time under intense radiation,
we developed new data acquisition (DAQ) system \cite{mizumoto_IEEE_2013},
which incorporates a new algorithm for electron track detection \cite{komura_IEEE_2013}. 
The TPC data are formatted in two ways.
The first is FADC data. 
The new readout board has 4~FADCs, and a time duration of 10~$\mu$sec is encoded to allow for the electron drift time in the TPC.
The other data format is the TOT data of all channels of the $\mu$PIC hit by the tracks.
The board has Boolean hit data of 128~channels $\times$ 1024 ~clocks (operating at 100~MHz for 10~$\mu$sec),
and the 30~cm$^3$~cubic ETCC has six readout boards.
The TOT data of a single event are typically sourced from 3~hit boards, 30~clocks and 30~strips.
However, SMILE-I\hspace{-.1em}I collects more TPC data per event than SMILE-I.
In all-transferred mode, all of these data are transferred at a fixed size of 27~kBytes.
Since the data transfer rate is restricted by the readout speed of VME (approximately 3~MByte/s),
the DAQ event rate is limited to below 110~Hz, whereas intense radiation experiments require rates of 500~Hz at least.
To solve this problem, we developed a data-suppressed mode.
The TPC data formats of all transferred and suppressed modes are shown in table \ref{tab:Trans_Data} and \ref{tab:Total_Data}.
In data-suppressed mode, the data size is reducible to 3.6~kByte on average by zero suppression,
erasure of the FADC data, and an efficient flag format.
Under these conditions, the DAQ event rate is limited at $\sim$800~Hz.

\begin{figure}[!h]
\centering
\includegraphics[width=.52\textwidth]{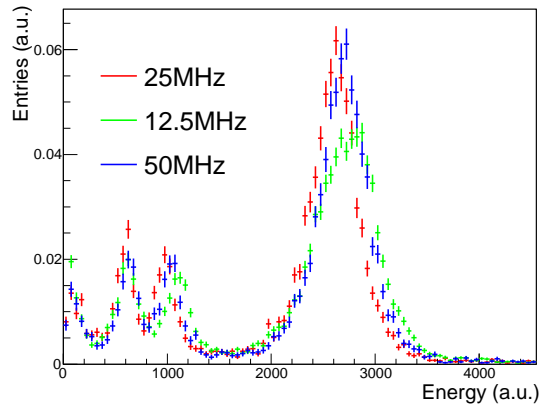}
\caption{
$^{133}$Ba spectra of (Anode, Cathode) = (10, 9) in the top panel of figure \protect\ref{fig:TPC}. \newline
Sampling rate are 25~MHz (red), 12.5~MHz (green) and 50~MHz (blue),
with energy resolutions of 20.1, 26.0, 19.2 $\%$ FWHM at 31~keV ($\sim$2700 a.u.), respectively.
}

\label{fig:Data_Reduction}
\end{figure}

Figure \ref{fig:Data_Reduction} shows the $^{133}$Ba spectra obtained under the altered FADC sampling rate.
At sampling rates of 25~MHz and 50~MHz, the energy resolutions are very similar (20~$\%$ FWHM),
but the resolution degrades to 26~$\%$ FWHM at 12.5~MHz.
Based on these spectra and the data size, we determined 25~MHz as an appropriate sampling rate in SMILE-I\hspace{-.1em}I.

\begin{table}[!h]
\caption{Transferred Data by one of $\mu$PIC Readout Boards}
\label{tab:Trans_Data}
\smallskip
\centering
\begin{tabular}{ | c || c  c | c c || }
\hline

& \multicolumn{2}{  c | }{FADC data}  & \multicolumn{2}{ c || }{TOT track data} \\
& all transferred & suppressed & all transferred & suppressed \\
\hline
Sampling rate [MHz]    & 50 (512 clocks) & 25 (256 clocks) & 100 & 100  \\
FADC                           & 10~bit $\times$ 4~ch &  8~bit $\times$ 4~ch  &  -     &  -      \\
Transferred channels   &  -   &  -   & 128 & 30 [typ.]    \\
TOT hit clock                &  -   &  -   & 30  [typ.] & 30 [typ.]    \\
Data transfer                & all 6~boards & only hit board & - & -  \\
\hline
Size of data                  & 2.5 kByte & 1~kByte   & 0.5~kByte  & 0.1~kByte  \\ 
                                     &     [fixed]  &     [typ.]     &  [typ.]         &      [typ.]      \\ \cline{2-5}
Size of other flag data  & 1.5 kByte &    0 kByte & 0.5~kByte  &  0.1~kByte  \\
                                     &     [fixed]  &                  &    [typ.]       &      [typ.]      \\

\hline
\end{tabular}
\end{table}

\begin{table}[!h]
\caption{Total size of TPC data}
\label{tab:Total_Data}
\smallskip
\centering
\begin{tabular}{ | c || c | c || c || }

\hline
& FADC & Track & Total \\
\hline
\hline
All-transferred mode
& 4 kByte $\times$ 6 boards & 1 kByte $\times$ 3 boards [typ.] & $\sim$27 kByte \\
\hline
Data-suppressed mode
& 1 kByte $\times$ 3 boards [typ.] & 0.2 kByte $\times$ 3 boards [typ.] & $\sim$3.6 kByte \\
\hline

\end{tabular}
\end{table}

\subsection{Field of View of ETCC}
\begin{figure}[htbp]
\centering
\includegraphics[width=.45\textwidth]{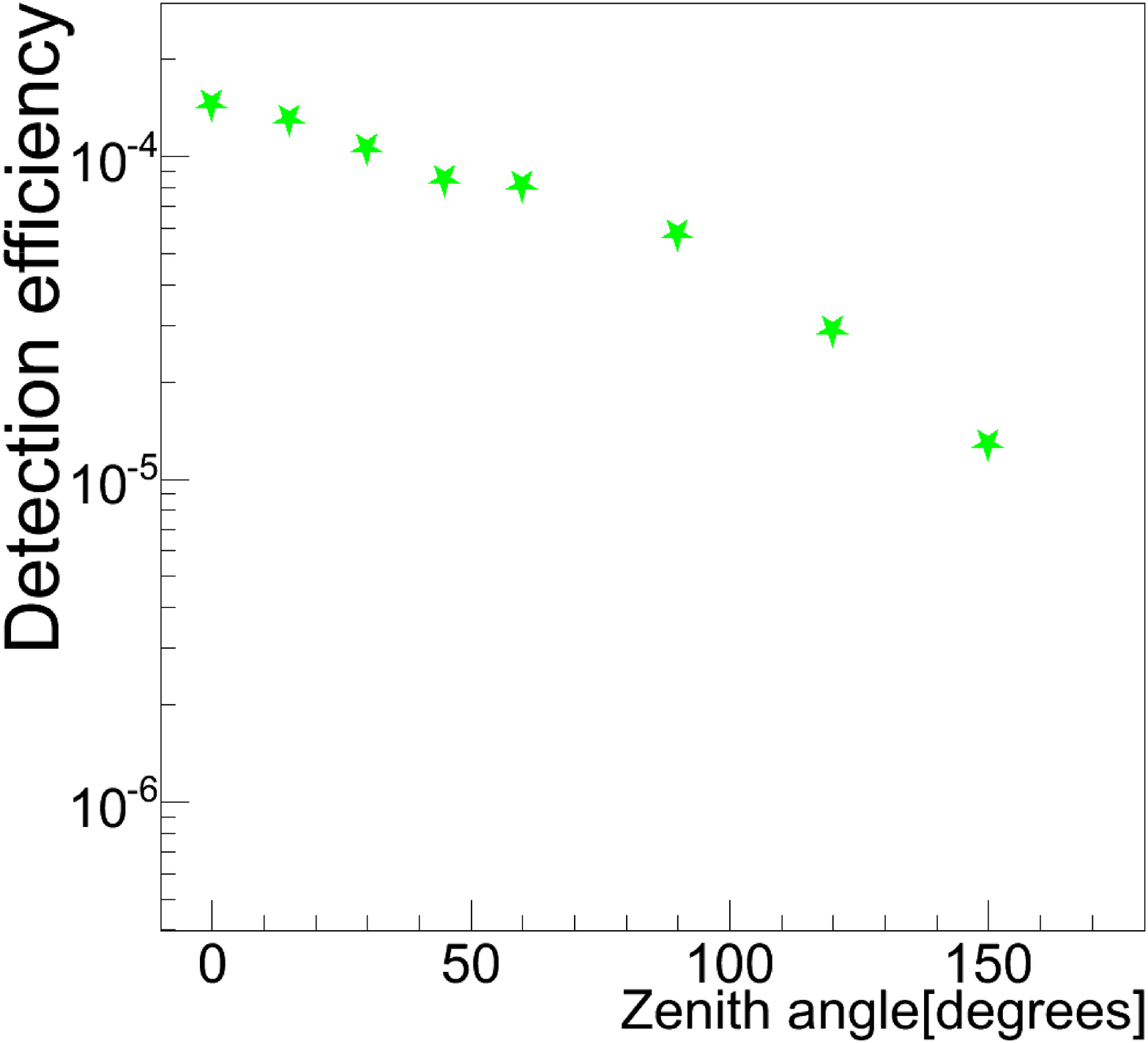}
\includegraphics[width=.5\textwidth]{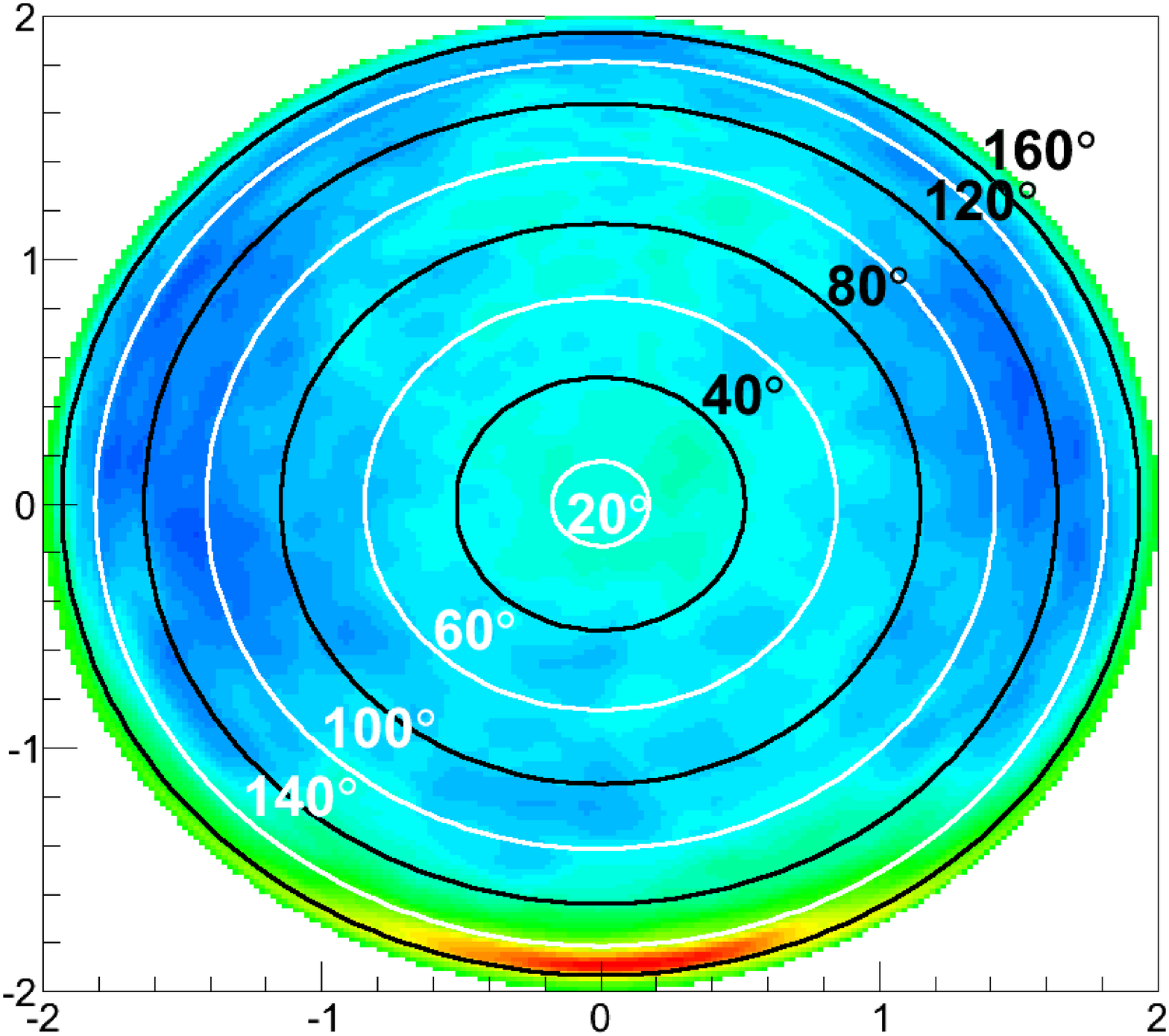}
\caption{
(\emph{Left panel}) Detection efficiency of ETCC at various zenith angles. The source is $^{137}$Cs (662~keV). \newline
(\emph{Right panel}) Image of the $^{137}$Cs source at 150$^{\circ}$ zenith.
}
\label{fig:Zenith}
\end{figure}

To detrain the available field of view (FoV) of the ETCC,
we measured the detection efficiency at various zenith angles with $^{137}$Cs (662~keV).
At 0$^{\circ}$ zenith, the ARM was measured as 5.3$^{\circ}$ FWHM \cite{mizumura_JINST_2014}.
The detection efficiency as a function of zenith angle is plotted in the left panel of figure \ref{fig:Zenith}. 
Since the efficiency at the FoV center falls by one-half at approximately 90$^{\circ}$,
the acceptable FoV extends to 2$\pi$~sr.
Additionally, the ETCC detects backward-incident gamma rays (right panel of figure \ref{fig:Zenith}),
indicating that our ETCC can reject background events such as atmosphere-sourced noise by image selection.

\section{Intense radiation experiment using proton beam}
\begin{figure}[htbp]
\centering
\includegraphics[height=5cm]{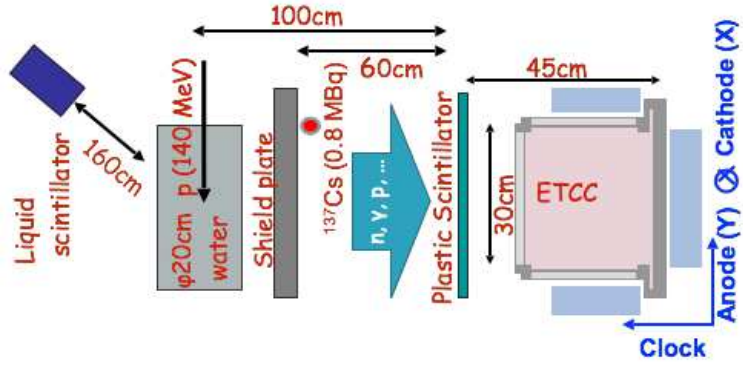}
\caption{
Setup of background rejection test.
}
\label{fig:RCNP_Setup}
\end{figure}

Observations at balloon altitudes or satellite orbits are obstructed by the prevailing huge background.
In particular, neutrons trigger observational instruments similarly to Compton scattering.
Moreover, neutrons are created when cosmic rays interact with the atmosphere or with the passive material around the detector.
These neutrons constitute most of the background of a Compton telescope.
However, because the proposed ETCC identifies particles by the dE/dX selection in the TPC,
it rejects such neutron-induced events.
To confirm the background rejection power of the ETCC,
we observed a faint gamma ray source with huge background generated by the 140~MeV proton beam
at the Research Center for Nuclear Physics, Osaka University.
The experimental setup is shown in figure \ref{fig:RCNP_Setup}.
The Bragg peak of the 140~MeV proton beam is approximately 14~cm in water, with a current of $\sim$0.1~nA on a water target.
The SMILE-I\hspace{-.1em}I ETCC was placed in the resulting strong radiation field composed of gamma rays and neutrons.

\begin{figure}[htbp]
\centering
\includegraphics[height=5cm]{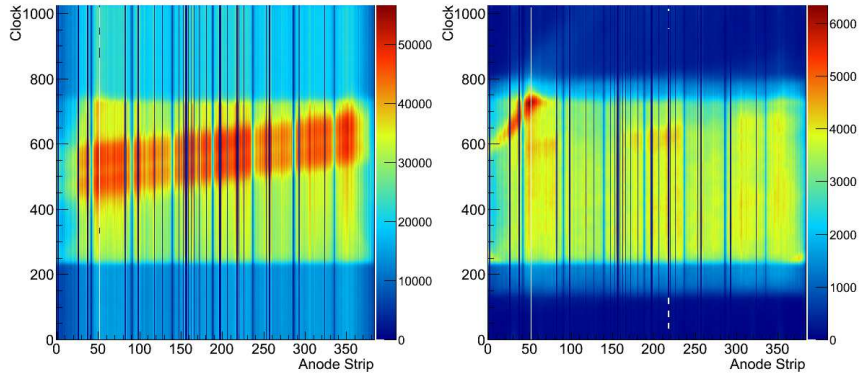}
\caption{
The left and right panels show histograms of tracks before and after dE/dX selection, respectively;
the x-axis (Anode Strip) indicates the horizontal direction parallel to the proton beam,
and the y-axis (Clock) indicates the depth (direction of the water target). \newline
The events at (Anode, Clock) = (50, 700) in the right are too short to be cut off by the dE/dX threshold.
}
\label{fig:RCNP_dEdX}
\end{figure}

From the ratio of neutrons to gamma rays (approximately 0.1) measured by a liquid scintillator,
the ratio entering the ETCC region was predicted as $\sim$1/3 by the Geant4 simulation toolkit \cite{ref_Geant4}.
To reduce the prompt gamma rays directly entering from the Bragg peak,
which would be registered as source rather than intense background,
the ETCC was placed perpendicular to the beam line 
and a 5~cm-thick lead shield was placed between the water target and the ETCC.
A checking source ($^{137}$Cs, 0.85~MBq) was placed behind the shield, 70~cm distant from the ETCC.
In data-suppressed mode, the SMILE-I\hspace{-.1em}I system acquired data at over 500~Hz,
five times faster than the estimated DAQ rate at mid-latitude balloon altitudea.
The panels of figure \ref{fig:RCNP_dEdX} are superimposed images of tracks detected by the TPC.
The diagonal lines in the center histogram are attributed to unexpected charged particle events striking directly from the proton beam line.
These background particles disappear after applying the dE/dX cutoff (right histogram of figure \ref{fig:RCNP_dEdX}).
This result indicates the powerful background rejection ability of ETCC enabled by particle identification.

\begin{figure}[htbp]
\centering
\includegraphics[height=5cm]{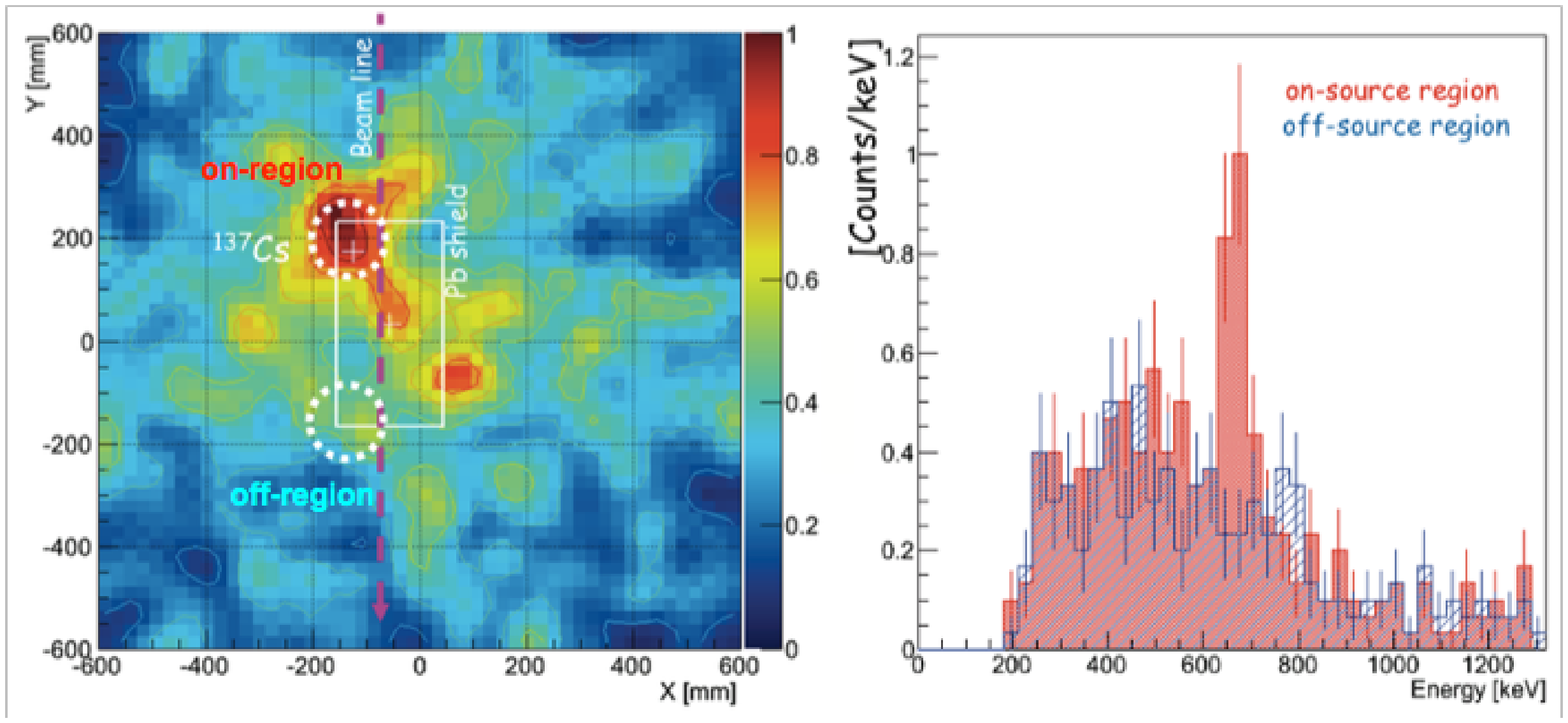}
\caption{
(\emph{Left panel}) Gamma ray image reconstructed from back projection in an energy window of 662~keV $\pm$ 10\%.
A clear peak appears at the checking source position. \newline
The events at (x = 100~mm, y = -100~mm) appear to originate from the point source,
but are more likely sourced from the stand holding the Pb shield. \newline
(\emph{Right panel}) Reconstructed gamma ray spectra in the on-source (red) and off-source (blue) regions.
The on-source region lies inside the upper dashed circle enclosing the gamma ray source position
(x = -124~mm, y = 200~mm, radius = 75~mm)
and the off-source region is symmetric about this position along the y-axis
(x = -124~mm, y = -200~mm, radius = 75~mm).
}
\label{fig:RCNP_source}
\end{figure}

The left panel of figure \ref{fig:RCNP_source} shows the reconstructed gamma ray image in the 662~keV~$\pm$~10$\%$ energy range.
The checking source is $^{137}$Cs.
In this image, the on-source region is defined to be inside the dashed circle around the source position (i.e., the upper dashed circle),
and the off-source region is assumed symmetric about this position along the y-axis.
The right panel of figure \ref{fig:RCNP_source} shows the spectrum in each region of the image.
The on-source spectrum shows a clear peak at 662 keV originating from the checking source,
which is absent in the off-source spectrum.
Meanwhile, the continuum components in the on-source and off-source spectra are consistent.
In this method, the checking source is detected to $\sim$5~$\sigma$ at a live time of $\sim$4500~s and a signal-to-noise ratio of $\sim$1/10.
Therefore, in the presence of huge background,
the source flux is accurately estimated by this simple on-off method.

\section{Summary}
We have developed an ETCC for use in MeV gamma ray astronomy. 
In particular, we constructed a mid-sized ETCC from a (30 cm)$^3$ TPC and 108 PSAs
for observing the Crab nebula in the next balloon flight (SMILE-I\hspace{-.1em}I).
The energy resolutions of the TPC and PSAs are 24.8~$\%$ FWHM at 31~keV and 11.1~$\pm$~0.6~$\%$ FWHM at 662~keV,
and the angular resolution of ETCC is 5.3$^{\circ}$ FWHM at 662~keV and 0$^{\circ}$ zenith.
The large FoV of the ETCC in SMILE-I\hspace{-.1em}I-FM ($2\pi$ sr) was also confirmed.
In data-suppressed mode, the ETCC can detect events at 500~Hz,
approximately five times the expected detection frequency at balloon altitudes.
Finally, the ETCC detected a $^{137}$Cs checking source to $\sim$5~$\sigma$
and demonstrated strong performance under huge background noise.
Therefore, without an active veto shielding, the ETCC has the large FoV and is effective for all-sky survey.

\acknowledgments
This study was supported by
a Grant-in-Aid for Scientific Research,
a Grant-in-Aid from the Global COE program ''The Next Generation of Physics, Spun from Universality and Emergence'' from the Ministry of Education, Culture, Sports, Science and Technology (MEXT) of Japan,
and the Japan Society for Promotion of Science for Young Scientists.
Also this study was supported by the joint research program of the Solar-Terrestrial Environment Laboratory, Nagoya University and the National Institute of Polar Research through General Collaboration Projects no 23-3.
Some of the electronics developments were supported by KEK-DTP and Open-It Consortium.

\end{document}